# Qualitatively Different Theoretical Predictions for Strong-Field Photoionization Rates

Jarosław H. Bauer *

*Katedra Fizyki Teoretycznej Uniwersytetu Łódzkiego,
Ul. Pomorska 149/153, PL-90-236 Łódź, Poland*

We give examples showing that two well-known versions of the $S$-matrix theory, which describes a nonresonant multiphoton ionization of atoms and ions in intense laser fields, lead to qualitatively different results. The latter refer not only to total ionization rates, but also to energy distributions of photoelectrons, for instance in a polarization plane of the laser field. It should be possible to make an experiment testing predictions of both theories in the near future.

___________________

*bauer@uni.lodz.pl

*Introduction.*-There are several methods, which describe the ionization of atoms and molecules in nonperturbative laser fields. Maybe recently the most popular and accurate method (called also an *ab initio* treatment) is numerical solving of the time-dependent Schrödinger equation (TDSE) for an initially bound electron (or electrons). However, considering computer limitations, older nonperturbative theories still have an advantage over the *ab initio* treatment, particularly for very strong laser fields of low frequencies. In the present work we demonstrate that two such approximate theories may lead to contradictory predictions in sufficiently intense fields. Moreover, the field parameters taken by way of example (see Figs. 2-5) indicate that future experiments might question one of these theories in some situations. The main aim of our work is to explicitly show these qualitative differences and to stimulate an experimental activity in this domain. We also hope that our results regarding photoelectron energy spectra might stimulate theoretical efforts having in view some simple mathematical formula describing positions of the maxima (see Figs. 4 and 5).

*S-matrix theories.* There are two well-known versions of the so-called strong-field approximation (SFA), which is the time-reversed *S*-matrix theory describing the nonresonant multiphoton ionization of atoms and ions in intense laser fields [1,2]. The main approximation utilized here is connected with a use of the Gordon-Volkov wave function [3,4], as a final state of an outgoing electron. Thus an effect of a binding potential on the escaping electron is completely neglected. In principle, one could think that this is a very crude approximation. However, the Gordon-Volkov wave function (instead of an exact solution to the TDSE) works very well, in the *S*-matrix theory, in the following two cases. The first one is connected with the zero-range model, which describes an outer (weekly bound) electron in the negative hydrogen ion placed in the nonperturbative laser field of an arbitrary polarization (see, for example, Refs. [5,6]). The second case is connected with the Coulomb potential and the sufficiently strong circularly polarized (CP) laser field [7-9]. In the latter case, due to selection rules, the ionized electron is forced to absorb on average much more photons (for a given electric field amplitude of the laser) than for a linearly polarized (LP) field. Already classical considerations [10,11] lead to a conclusion that in the intense CP laser field ionized electrons should have a kinetic energy peak near its ponderomotive energy $U_P$ (sometimes called the ponderomotive potential as well) of the interaction of a free electron with the field. (For the LP laser field ionized electrons always have the kinetic energy peak not far from zero [1,2]). Since for strong fields $U_P \gg E_B$ (the binding energy), an effect of the Coulomb potential on the final state of the outgoing electron should indeed be very small. [In the present work we use atomic units ($a.u.$): $\hbar = e = m_e = 1$, and we substitute explicitly $-1$ for the electronic charge.]

The basic difference between the two pioneering works of Keldysh [1] and Reiss [2] is the Hamiltonian form of the laser-atom interaction used to evaluate an amplitude of the ionization probability. Keldysh used the Hamiltonian in the length gauge (LG), while Reiss used this Hamiltonian in the velocity gauge (VG). The common feature of both approaches [1,2] (but also present in numerous later papers) was an application of nonrelativistic and dipole approximations to a description of this interaction. However, the work of Keldysh [1] concerned only the LP field of a low frequency. Keldysh made some further assumptions (which are absent in Ref. [2]) to get his final analytical results. In the present Letter we would like to focus on predictions of the $S$-matrix theory in both gauges only for the CP field, because for the LP field such predictions are not so much different [12]. In the former case the LG counterpart of the VG SFA of Reiss [2,7] is Ref. [9], where no further Keldysh-type assumptions have been done. The LG SFA from Ref. [9] and the VG SFA from Refs. [2,7] are physically equivalent (see Sec. III of Ref. [13]). Moreover, in Ref. [9] the LG SFA has been extended to initial states (of the hydrogenic atom) with the principal quantum number $n = 2$. In Refs. [2,7,9] one assumes that two main conditions, among others, are satisfied in the nonrelativistic $S$-matrix theory:

$$z_1 \equiv \frac{2U_P}{E_B} = \frac{I}{2\omega^2 E_B} \gg 1 \; , \tag{1}$$

and

$$z_f \equiv \frac{2U_P}{c^2} = \frac{I}{2\omega^2 c^2} \ll 1 \; , \tag{2}$$

where $I$ stands for an intensity of the laser field and $c$ is the speed of light. The parameters $z_1$ and $z_f$ have been introduced by Reiss [2]. For the CP field $z_1 = 2/\gamma^2$, where $\gamma$ stands for a well-known Keldysh adiabaticity parameter [1]. A qualitative difference between the VG SFA and the LG SFA appears already in general expressions describing their respective ionization probability amplitudes, namely

$$(S-1)_{fi}^{VG\ SFA} = i \int_{-\infty}^{\infty} dt \widetilde{\Phi}_i(\vec{p}) \left( \frac{1}{2}\vec{p}^{\,2} + E_B \right) \exp\left[ \frac{i}{2} \int_{-\infty}^{t} \left( \vec{p} + \frac{1}{c}\vec{A}(\tau) \right)^2 d\tau + iE_B t \right], \qquad (3)$$

$$(S-1)_{fi}^{LG\ SFA} = i \int_{-\infty}^{\infty} dt \widetilde{\Phi}_i\left( \vec{p} + \frac{1}{c}\vec{A}(t) \right) \left( \frac{1}{2}\left( \vec{p} + \frac{1}{c}\vec{A}(t) \right)^2 + E_B \right) \exp\left[ \frac{i}{2} \int_{-\infty}^{t} \left( \vec{p} + \frac{1}{c}\vec{A}(\tau) \right)^2 d\tau + iE_B t \right],$$

(4)

where $\vec{p}$ is the asymptotic momentum of the ionized electron, $\widetilde{\Phi}_i(\vec{p})$ is the initial-state wave function in the momentum representation, and $\vec{A}(t)$ is the vector potential of the laser field. (For a derivation of these formulas and ionization rates, and for more detail see Sec. III-V of Ref. [13].) The electric field component of the laser is present in Eqs. (3) and (4) through the relation $\vec{F}(t) = (-1/c)\partial \vec{A}(t)/\partial t$, and there is no magnetic field component of the laser here. Let us note that only for a specific choice of the initial-state wave function these two amplitudes (Eqs. (3) and (4)) become identical. This is a well-known case of the zero-range binding potential [5,6], when $\widetilde{\Phi}_i(\vec{p}) \sim (\vec{p}^{\,2}/2 + E_B)^{-1}$. For all other binding potentials, including the Coulomb one, both amplitudes have to differ as a matter of fact. In the VG SFA the product $\widetilde{\Phi}_i(\vec{p})(\vec{p}^{\,2}/2 + E_B)$ does not depend on time and can be taken in front of the integral in Eq. (2), what leads to an analytical simplicity in further calculations. However, there is also a very serious drawback of the VG SFA. As it has been shown recently both for the LP [14] and for the CP [15] laser fields, the VG SFA ionization rates vanish in the quasistatic limit [i.e. when (the field amplitude) $F = const$ and (the laser frequency) $\omega \to 0$] for the Coulomb potential. (In both works [14,15] nonrelativistic and dipole approximations have been applied.) This result is apparently unphysical, because in the quasistatic limit the ionization is caused only by a static electric field (for the CP field) or by the static electric field averaged over a laser period $T = 2\pi/\omega$ (for the LP field) [16,17]. Therefore such ionization rates should be nonzero and should depend on $F$, on $E_B$, and on the wave function $\widetilde{\Phi}_i(\vec{p})$. (For more detail see Sec. II of Ref. [13].) One should stress again that the absence of the ionization in the VG SFA theory (for the Coulomb potential) in the quasistatic limit is in an apparent contradiction with both theories and experiments describing atoms in constant electric (not laser) fields.

*Ionization rates.*-On the other hand, let us note that non-vanishing ionization rates (in the quasistatic limit) in the CP field exist when $\tilde{\Phi}_i(\vec{p}) \sim (\vec{p}^2/2 + E_B)^{-1}$, i.e. when both amplitudes (3) and (4) are equal. Then the respective, well-known, asymptotic expression is $\Gamma = (F/2\kappa)\exp(-2\kappa^3/3F)$ (with $E_B = \kappa^2/2$) (see Eq. (3.7) in Ref. [5], Eq. (22) in Ref. [15], and Ref. [18]). This expression becomes exact in the limit $F \to 0$. A more accurate expression for the ionization rate depends (through $\gamma$) also on $\omega$ [5,15,18]

$$\Gamma_{zero-range}^{asympt} = \frac{F}{2\kappa}\exp\left[-\frac{2\kappa^3}{3F}\left(1-\frac{\gamma^2}{15}\right)\right]. \tag{5}$$

The VG SFA ionization rate (which is the same as in the LG SFA) is well-known (Ref. [2] or Eq.(21) in Ref. [15]). One usually shows functions $\Gamma(F)$ (or $\Gamma(I)$; $I = 2F^2$ for the CP field) for some $\omega = const$. But it is very instructive to look at the function $\Gamma(\omega)$ for some $F = const$. We have done this in Figs. 1-3 in a total applicability range of the nonrelativistic SFA theory. For each curve shown here the lowest frequencies correspond to $z_f = 0.1$ and the highest ones – to $z_1 = 1$. The word "exact" in Figs. 1-3 means that the ionization rates have been computed using respective expressions derived long ago by Reiss in Ref. [2] or recently by us (for the LG) in Ref. [9]. The word "asymptotic" in Figs. 1-3 means that the ionization rates have been computed using Eqs. (22) or (13) (valid in the limit $F = const$ and $\omega \to 0$), respectively, from Ref. [15]. A derivation of asymptotic expressions seems to be a rather difficult task for the LG SFA and the Coulomb potential.

In Fig. 1 we present the SFA ionization rates for the zero-range binding potential with $E_B$ which corresponds to a negative hydrogen ion $H^-$ and two different values of the electric field. Fig. 1 is an example demonstrating that gauge-invariant ionization rates always approach some constant (positive) value when $F = const$ and $\omega \to 0$. The situation changes qualitatively when one considers the Coulomb binding potential instead of the zero-range binding potential. This is illustrated in Figs. 2 and 3. The LG SFA ionization rates still approach some positive value when $F = const$ and $\omega \to 0$, but the VG SFA ionization rates do not. They approach zero, as we have noticed above. Asymptotic expressions in the VG SFA behave as $\sim \omega^4$ for the $H(1s)$ atom. The power of $\omega$ describes the slope of slanted lines in the log-log plots from Figs. 2 and 3. The same behavior in this gauge is shown for the

"exact" rates when $\omega \ll E_B$. For an arbitrary $F$ and a sufficiently low $\omega$ one can always find a situation when the VG SFA ionization rate is many orders of magnitude smaller than its LG SFA counterpart. This could be tested experimentally for any atom.

The LG SFA ionization rates are not accurate in the quasistatic limit, because to obtain them one neglects the (long-range) Coulomb potential in the final state of the ionized electron. For $F = 1$ *a.u.* the agreement with the accurate $\omega = 0$ results is better than for $F = 0.05$ *a.u.*, because Coulomb effects are relatively weaker in the former case. Therefore, these rates can be even with an order-of-magnitude agreement with the accurate $\omega = 0$ results. The latter are numerical results of Scrinzi *et al* [19,20] (we show them for a comparison in Figs. 2 and 3 as well) for the $H(1s)$ atom in the static nonperturbative electric field. Also the experimental data of Buerke and Meyerhofer [21] for the low-frequency ($\omega \approx 0.043$ *a.u.*) ionization of the $He^+(1s)$ ion in the CP laser field are in a better agreement with the LG SFA ionization rates. We discuss this fact in more detail late in Ref. [22]. Taking into consideration some Coulomb correction in the final state of the ionized electron one may significantly improve the VG SFA theory, but it is still worse than its LG counterpart [22].

It has been argued recently [23] that in the quasistatic limit (then also $\gamma \to 0$ for any $F = const$) the laser field becomes so strong (in a sense that then Eq. (1) is amply satisfied and Eq. (2) violated, because $z_1 \to \infty$ and $z_f \to \infty$) that the dipole approximation breaks down. Indeed, in superstrong laser fields first nondipole (i.e. connected with a magnetic-field component of an electromagnetic plane wave) and then relativistic effects have to be taken into account [24]. However, the magnetic-field component of the strong, but nonrelativistic, laser field is less essential in the CP field than in the LP field [25,26]. In the CP field, in the simplest frame of reference (see [25,26] and references therein) a charge in the plane-wave laser field always moves along a circle lying in the polarization plane. Relativistic effects in the VG SFA theory (within the Dirac formalism for the $H(1s)$ atom in the CP laser field) were studied in Ref. [8]. It was shown that for sufficiently intense laser fields (roughly speaking, when $z_f > 1$) relativistic ionization rates are even smaller than nonrelativistic ones (see Figs. 1 and 2 in Ref. [8]). Therefore, it is likely that such relativistic ionization rates approach zero even faster (when $F = const$ and $\omega \to 0$) than their nonrelativistic counterpatrs. As a result, taking into account relativistic effects in the VG SFA theory does not solve the problem of vanishing rates. Instead of this behavior, ionization rates should approach (nonzero) values, which describe the ionization in constant perpendicular electric and magnetic fields of the same magnitude (in atomic units; see, for example, Ref. [24]).

*Energy spectra.*-Another physical quantities, which show qualitative differences between the VG SFA and the LG SFA, are probability distributions of ionized electrons (photoelectron energy spectra). In Ref. [9] we have found such remarkable differences for the ionization in intense CP laser fields of the hydrogen atom in some initial states with a principal quantum number $n = 2$. In an experiment one usually prepares an atom in the initial state with a single set of the $(n,l,m)$ quantum numbers (we omit spin effects here). Therefore, we have generalized the LG SFA theory (from Ref. [9]) to initial states $(2,1,-1)$ and $(2,1,1)$ [12]. In Figs. 4 and 5 we present energy spectra of photoelectrons (i.e. differential ionization rates: $\partial^2 \Gamma / \partial E \partial \vartheta$) in the polarization plane ($\vartheta = \pi/2$) for the intense ($z_1 = 1000$) CP laser field. Since the ionized electrons are emitted mostly in this plane one obtains similar pictures for the spectra ($\partial \Gamma / \partial E$) integrated over a full solid angle. The LG SFA ionization rates are usually a few orders of magnitude larger than their VG SFA counterparts. In Figs. 4 and 5 we compare shapes of two probability distributions. To this end the VG SFA differential ionization rates have been multiplied by a suitable factor (much larger than 1) to get the same area under both curves (vertical axes have a linear scale in Figs. 4 and 5). There is also a vertical dashed line, which displays the ponderomotive energy $U_P$. The VG SFA energy distributions have a single-peak shape. The peak is near $U_P$, but with a little shift towards lower energies. These energy distributions are identical for both initial states: $(2,1,-1)$ and $(2,1,1)$. The LG SFA energy distributions, which show a double-peak shape, are qualitatively different. They have peaks situated on both sides of the vertical line $E = U_P$. Moreover, the relative height of these peaks changes rapidly, if one changes the azimuthal quantum number from $m = -1$ to $m = 1$ (or reversibly). There is a minimum between both peaks at $E \approx U_P$. Positions and heights of these two peaks are not clear to us, because our analytical calculations do not allow for a simple interpretation. We have not been able to find any simple mathematical formula even for positions of these two peaks. However, a nature of the peaks may be connected with a fact that the initial states $(2,1,-1)$ and $(2,1,1)$ have the $z$-component of an angular momentum parallel or antiparallel with respect to the angular momentum carried by the laser field (which propagates along the $z$ axis). On the other hand, the initial state $(2,1,0)$ also shows a multipeak spectrum in the LG SFA [9,12]. A similar effect was observed quite long ago in the ionization of the same atom in a weaker 20-cycles sine-square laser pulse (with $\omega = 2E_B$) by Gajda *et al.* [27]. In this *ab initio* calculations the asymmetry between the initial states $(2,1,-1)$ and $(2,1,1)$ appears in photoelectron energy spectra after switching-off the laser field. Therefore the results of Ref. [27] are gauge-invariant. It would be very interesting to compare predictions of the present LG SFA theory with the data of Gajda *et al.* (by integrating our ionization rates over the pulse profile). Moreover, since no dependence on the azimuthal quantum number (in the initial state with $n=2$) is present in the VG SFA, this theory and the results of Ref. [27] contradict each other.

*Final remarks.*-We would like to mention about yet another qualitative difference between the LG and the VG $S$-matrix theories, which has been already verified experimentally. Bashkansky *et al.* [28] have shown a violation of a fourfold symmetry (to a two-fold one) in photoelectron angular distributions generated by an elliptically polarized light (using helium, krypton and xenon atoms). The fourfold symmetry is predicted, if one applies the Gordon-Volkov wave function as the final state of the outgoing electron. Subsequent theoretical explanations [29,30] have shown that it is necessary to use the LG (but not the VG) together with some Coulomb-Volkov wave function (in the final state) to recover qualitatively the results of Bashkansky *et al.* On the other hand, if one includes the laser field dressing in the initial state and takes the final state as a pure Coulomb scattering state, one can also get the proper symmetry in angular distributions in the VG [31].

In our opinion, there are deeper reasons for qualitative differences, which characterize both versions of the SFA theory. It has been observed quite long ago that the VG SFA is "a hybrid procedure" from the point of view of a gauge consistency [32]. In the VG SFA one uses eigenstates of an "unperturbed" Hamiltonian $H_0 = -\Delta/2 + V(\vec{r})$ as reference states. Contrary to that, in the LG SFA one employs eigenstates of the physical energy operator, which is gauge-invariant (or gauge-covariant) [33-35] (for more detail see references therein).

I thank Howard R. Reiss for his correspondence on subjects related to this work.

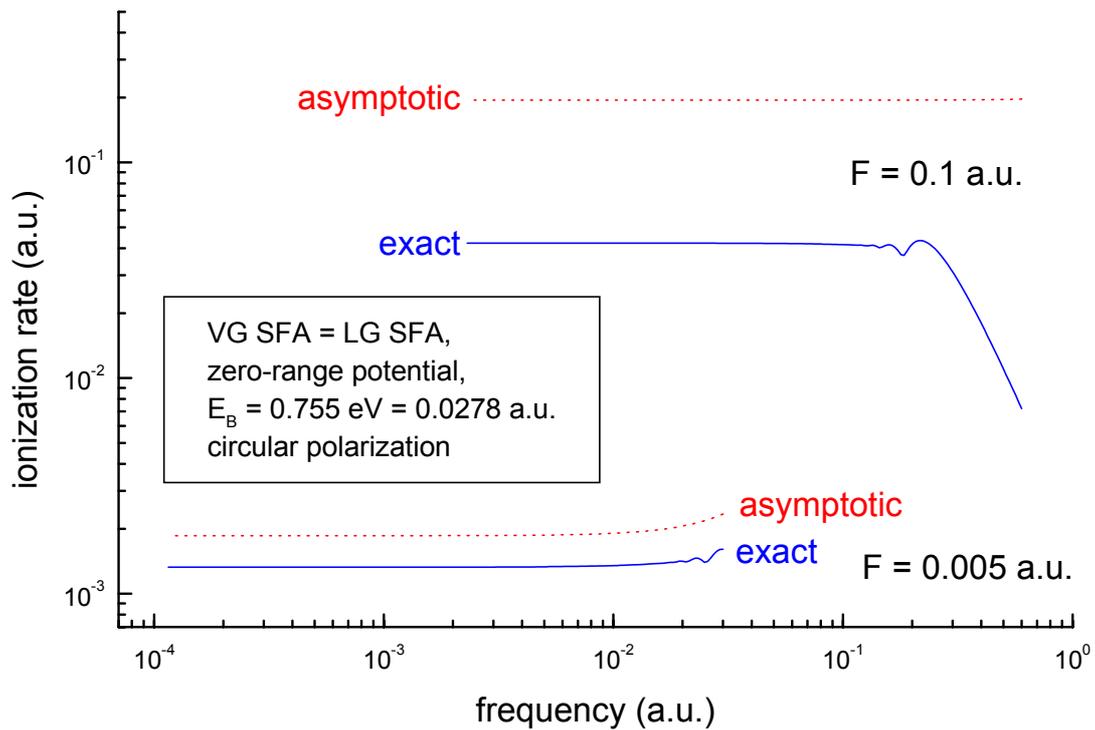

FIG. 1. (Color online) The gauge-invariant SFA ionization rates of the $H^-$ ion for $F = 0.1$ *a.u.* and $F = 0.005$ *a.u.* vs. ω (see the text for more detail).

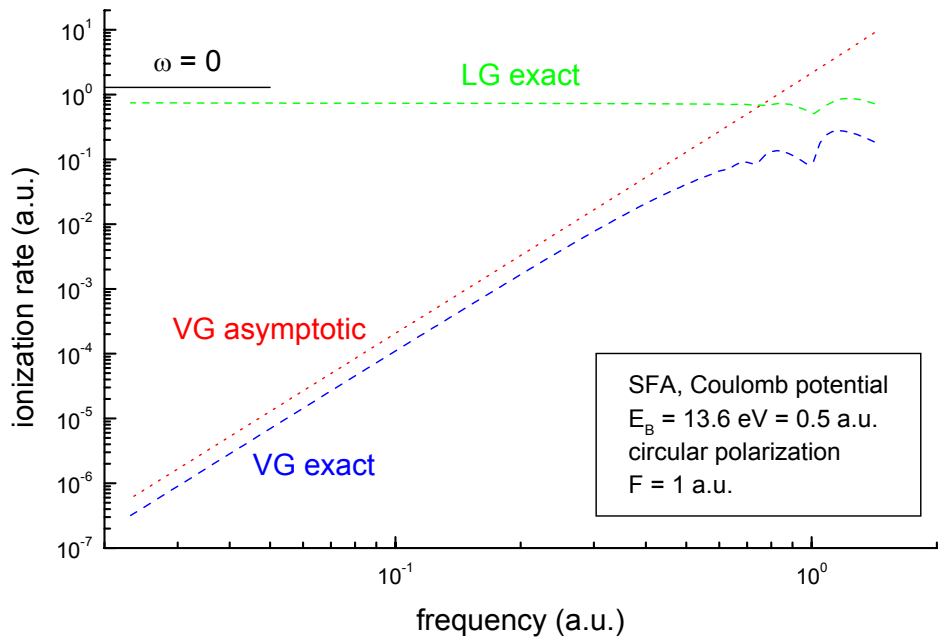

FIG. 2. (Color online) The LG SFA and the VG SFA ionization rates of the $H(1s)$ atom for $F = 1$ $a.u.$ vs. $\omega$. The exact static-field result of Scrinzi et al. [19,20] ($\omega = 0$) is also shown for a comparison (see the text for more detail).

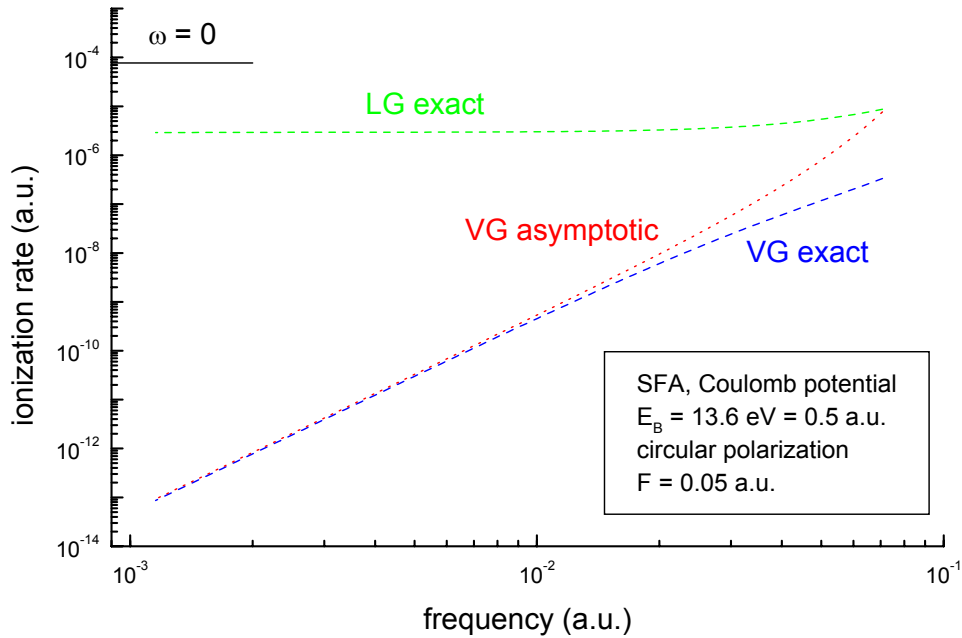

FIG. 3. (Color online) As Fig. 2, but for $F = 0.05$ $a.u.$

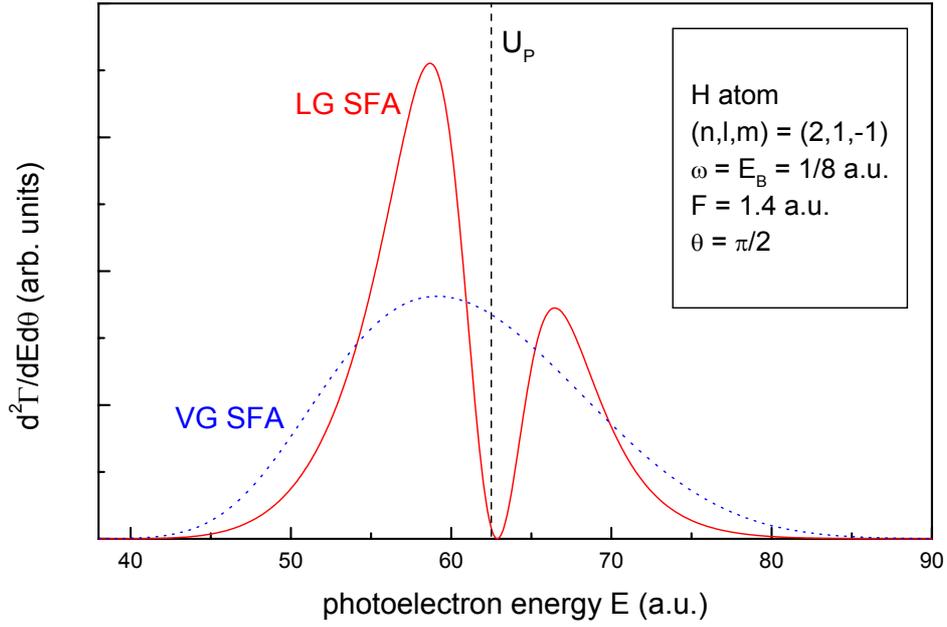

FIG. 4. (Color online) The LG SFA and the VG SFA differential ionization rates of the $H(2p)$ atom in the polarization plane ($\vartheta = \pi/2$) of the CP laser field. The azimuthal quantum number is $m = -1$, $F = 1.4$ $a.u.$ (what corresponds to $z_1 = 1000$ or $\gamma = 0.045$), and $\omega = E_B = 0.125$ $a.u.$ (see the text for more detail).

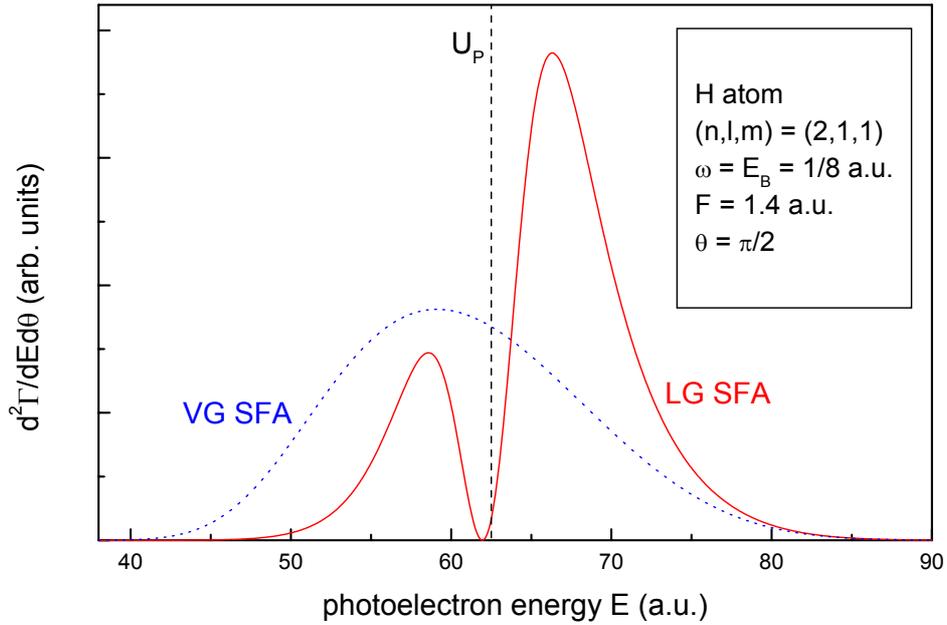

FIG. 5. (Color online) As Fig. 4, but for $m = 1$.